\documentclass[conference]{IEEEtran}
\IEEEoverridecommandlockouts
\usepackage{cite}
\usepackage{amsmath,amssymb,amsfonts}
\usepackage{algorithmic}
\usepackage{graphicx}
\usepackage{textcomp}
\usepackage{xcolor}

\usepackage{times}

\usepackage{soul}
\usepackage{url}
\usepackage[hidelinks]{hyperref}
\usepackage[utf8]{inputenc}
\usepackage[small]{caption}
\usepackage{graphicx}
\graphicspath{ {./images/} }
\usepackage{amsmath}
\usepackage{amsthm}
\usepackage{booktabs}
\usepackage{amssymb}
\usepackage{latexsym} 
\usepackage{threeparttable}
\usepackage{subfigure}
\usepackage{multirow}

\def\BibTeX{{\rm B\kern-.05em{\sc i\kern-.025em b}\kern-.08em
    T\kern-.1667em\lower.7ex\hbox{E}\kern-.125emX}}
\begin{document}

\title{
Scenario-aware and Mutual-based approach for Multi-scenario Recommendation in E-Commerce 
}



\author{
	\IEEEauthorblockN{
		Yuting Chen$^{1,2*}$
		\thanks{$^*$Both authors contributed equally to this research.
		        Work done when Yuting Chen was intern at Alibaba Group.},
		Yanshi Wang$^{2*}$, 
		Yabo Ni$^{2}$, An-Xiang Zeng$^{2}$, Lanfen Lin$^{1}$ 
	}

	\IEEEauthorblockA{
		{$^1$Zhejiang University, Hangzhou, China} \\
		{$^2$Alibaba Group, Hangzhou, China} \\
	 		21851477@zju.edu.cn,
	 		\{{yanshi.wys, yabo.nyb\}@alibaba-inc.com,
	 		renzhong@taobao.com,
	 		llf@zju.edu.cn}
	}
}

\maketitle

\begin{abstract}

Recommender systems (RSs) are essential for e-commerce platforms to help meet the enormous needs of users.
How to capture user interests and make accurate recommendations for users in heterogeneous e-commerce scenarios is still a continuous research topic. 
However, most existing studies overlook the intrinsic association of the scenarios: 
the log data collected from platforms can be naturally divided into different scenarios (e.g., country, city, culture). 
We observed that the scenarios are heterogeneous because of the huge differences among them.
Therefore, a unified model is difficult to effectively capture complex correlations (e.g., differences and similarities) between multiple scenarios thus seriously reducing the accuracy of recommendation results.

In this paper, we target the problem of multi-scenario recommendation in e-commerce, and propose a novel recommendation model named Scenario-aware Mutual Learning (SAML) that leverages the differences and similarities between multiple scenarios. 
We first introduce scenario-aware feature representation,
which transforms the embedding and attention modules to map  the features into both global and scenario-specific subspace  in parallel.
Then we introduce an auxiliary network to model the shared knowledge across all scenarios, and use a multi-branch network to model differences among specific scenarios. 
Finally, we employ a novel mutual unit to adaptively learn the similarity between various scenarios and incorporate it into multi-branch network.
We conduct extensive experiments on both public and industrial datasets, empirical results show that SAML consistently and significantly outperforms state-of-the-art methods. 
\end{abstract}


\section{Introduction}
In the Internet era, large global e-commerce portals such as Amazon and AliExpress often serve customers all over the world and contain billions of  items. 
It is particularly challenging for the recommender systems to meet the enormous needs of users with different preferences.
Personalization techniques are critical for these systems in that modeling user's interests more precisely can help improve user experience and generate more business value.

In practice, the log data collected from e-commerce portals can be naturally divided into different scenarios (e.g., country, city, culture).
Those scenarios are heterogeneous and there may be complex correlations between different scenarios, such as huge differences in user's interests, preferences, etc. On the contrary, in some cases users may have some similar interests as well (like countries with similar geographic locations).

There has been a large body of researches in recommender systems~\cite{zhang2017joint, lian2018towards}.
Most of the researches are based on deep neural networks (DNNs) and recurrent neural networks (RNNs). 
More recently, attention mechanism~\cite{zhou2018deep, zhou2019deep, feng2019deep, chen2019behavior} has been introduced for better performance. Many of these techniques have been successfully deployed in real-world applications~\cite{he2014practical, covington2016deep, borisyuk2017lijar}. 
However, existing recommendation methods mainly ignore the complex correlations between multiple scenarios and simply apply a general model for all scenarios, which may be sub-optimal as valuable information is not clearly captured across different scenarios.

Regarding the above issues, the intuitive consideration is to model for each scenario with the data of itself respectively. However, this may cause insufficient training problems on part of small-traffic scenarios and ignore the correlation between scenarios as well. 
In addition, Multi-Task Learning (MTL) may be a feasible solution. By treating each scenario as a separate task, a MTL scheme can be used to model the correlation between multiple scenarios, such as  MMoE~\cite{ma2018modeling} that implicitly integrate information between relevant scenarios through MoE structure and gate units.

Different from existing research, this work focus on perceiving scenario awareness in an explicit manner. 
We propose \textbf{S}cenario-\textbf{a}ware \textbf{M}utual \textbf{L}earning (SAML) which targets at learning both global and scenario-specific representations across multiple scenarios simultaneously. 
The global representation can extract shared knowledge from various scenarios, and the scenario-specific representation can learn the specific representation of each scenario individually.



In practice, we first build both global and scenario-specific subspace for embedding and attention module, and then combine the features from each subspace respectively to construct two types of features, named scenario-independent and scenario-dependent features.
Second, an auxiliary network and a multi-branch network are established upon the features to learn shared knowledge across scenarios as well as scenario-specific representations respectively.
Finally, a novel mutual unit is designed and incorporated into multi-branch network to capture correlations among multiple scenarios, which not only maintain the dominance of the current scenario but also leverages the knowledge from some of the similar scenarios adaptively.

Extensive experiments have been conducted to verify the effectiveness of the scenario-aware mutual learning (SAML) method. Evaluations of Click-Through Rate (CTR) prediction on public and industrial datasets show that the proposed SAML can generate better performance for multi-scenario recommendation compared to most advanced recommendation methods (without an explicit perception of scenario). 
Furthermore, ablation studies and visualization analysis on real-world industrial datasets demonstrate the proposed SAML does have an effective generalization. 


The main contributions of this paper are summarized as follows:
\begin{itemize}
    \item Considering the feature diversity in multiple scenarios, we transform the embedding and attention module to map the features into both global and scenario-specific subspace, and then combine the feature vectors in the corresponding subspace to construct the scenario-independent and scenario-dependent features respectively.
    \item We propose to learn scenario-independent and scenario-dependent features separately, thus an auxiliary network, as well as a multi-branch network, are built to learn deep representations from corresponding feature spaces respectively.
    \item We propose to model the complex correlations between multiple scenarios in an explicit manner, thus introduce a mutual unit and incorporate it into the multi-branch network to simultaneously model the differences and similarities between scenarios, which can maintain the dominance of the current scenario and leverage the knowledge from some of the similar scenarios adaptively.
    \item We conduct extensive experiments on both public and industrial datasets. Experimental results show that our proposed SAML can generate more accurate results in the multi-scenario recommendation task, and can also generalize to other scenarios effectively. 
\end{itemize}

The remaining parts of this paper are organized as follows: Section 2 introduces some related work. Section 3 describes and analyses the design of proposed SAML in detail. Experimental results and corresponding analysis are presented in Section 4 and conclusion in Section 5.

\section{Related Work}
In this section, we mainly introduce existing studies of feature representation and  multi-task learning in recommender systems as well as deep mutual learning.    
    \subsection{Feature Representation in Recommendation}
    In recommender systems, feature representation plays an important role in estimating the probability of corresponding user events, e.g., click and purchase. Enormous efforts have been put into modeling efficient features interactions and varied sequential behaviors. 
    
    Wide\&Deep~\cite{cheng2016wide} combines the benefits of linear and deep representations, serving as a good solution for this task.
    DeepFM~\cite{guo2017deepfm} replaces the wide component of Wide\&Deep with factorization machines (FM) to model second-order feature interactions.
    DCN~\cite{wang2017deep} further introduces a multi-layer residual~\cite{he2016deep} structure to learn high-order representation of features.
    
    Besides, users' sequential behavior implies the dynamic and evolving interests and has been proven effective in tasks of user interest estimation. DIN~\cite{zhou2018deep} applies attention mechanism to learn the representation of users' historical behaviors towards the target item.
    DIEN~\cite{zhou2019deep} further introduces an auxiliary loss and AUGRU to capture the evolution trend of users' interests.
    DSIN~\cite{feng2019deep} divides users' behaviors into different sessions and use self-attention to extract users' interests in each session. Most recently, BST~\cite{vaswani2017attention} deploys Transformer to the E-commerce recommendation and verify its effectiveness.
    
    However, these existing methods are mainly designed without consideration of multiple scenarios, thus the learned feature representation is from a global perspective and only in a homogeneous representation space, e.g., learning unified feature embedding and attention vector to express feature attributes and users' interests across various situations. This may become a bottleneck for distinguishing the interests of different users in multiple (heterogeneous) scenarios. 
    
    \subsection{MTL in Recommendation}
    MTL~\cite{caruana1997multitask, argyriou2008spectral} have been actively researched in recommender systems and numerous deep learning applications benefit from the multi-objective optimization. 
    DUPN~\cite{ni2018perceive} proposes a robust and practical representation learning framework, which learns sharing user representations in an end-to-end setting across multiple e-commerce tasks. 
    Considering the sequential pattern of user actions, ESMM~\cite{ma2018entire} introduces two auxiliary networks for CTR and CTCVR tasks, tackling the challenges of \textit{sample selection bias} and \textit{data sparsity} problems. 
    ESM$^2$~\cite{wen2019conversion} further decomposes the post-click behavior for modeling CVR task in e-commerce recommender system.
    MMoE~\cite{ma2018modeling} uses computational efficient Mixture-of-Experts (MoE)~\cite{jacobs1991adaptive} as shared-bottom as well as light-weight gating network to model task relationship, which proved can better handle the scenario where tasks are less related.
    
    In the context of our problem, although we can build individual networks for each scenario on top of a shared-bottom structure, and do  multi-objective optimization as classical MTL methodology, thereby modeling complex correlations between multiple scenarios. 
    However, when scenarios in recommender system share the same item candidates and label space, the consistency and discrepancy of scenarios are coupled with each other tightly, thus the sophisticated relationship between different scenarios are hard to capture. 
    
    \subsection{Deep Mutual Learning}
    Deep mutual learning~\cite{zhang2018deep} is proposed for knowledge distillation~\cite{hinton2015distilling}, which builds an ensemble of student networks to teach each other with distillation losses of Kullback-Leibler divergence. Inspired by this learning strategy, we explore a different idea to solve multi-scenario problem with a novel mutual unit to learn collaboratively scenario correlations in recommender systems. The essential difference is that our dataset shares a consistent label domain and the mutual unit uses a tailored similarity and gate mechanism to control the learning process instead of indirect approximation of data distribution by distillation. 

\section{Methods}
    In this section, we elaborate on the design of Scenario-aware Mutual Learning (SAML) model. First, we recapitulate the basic structure of deep learning based recommendation model from two aspects: feature representation and multi-layer perceptron. And then we introduce the overall structure of SAML corresponding to the above two aspects respectively.
    
    \subsection{Feature Representation}
    \label{sec 3.1}
    There are four categories of features in our recommender system: \textit{User Profile}, \textit{Item Profile}, \textit{User Behavior} and \textit{Context}. Each category of feature has several fields, \textit{User Profile} contains \textit{user\_id}, \textit{gender}, \textit{age} etc.; \textit{Item Profile} contains \textit{item\_id}, \textit{shop\_id}, \textit{price}, etc.; \textit{User Behavior} is the sequential list of user behavior, which contains the user's interacted items with corresponding features such as \textit{item\_id}, \textit{shop\_id}, etc.; \textit{Context} contains \textit{time}, \textit{matchtype}, \textit{scenario} and so on. 
    
    \subsubsection{Embedding Module}
    Features in each field are numerical value or categorical id. For numerical features, we use normalization to transform them into the same scale. For categorical features, which typically represented by one-hot vectors, we use embedding technology to transform them into low-dimension dense vectors. For example, the embedding matrix of \textit{item\_id} can be represented by $E_{item} = [e_1;e_2;...;e_K] \in R^{N \times K}$, 
    where $N$ is the total number of different items, $K$ is the dimension size of the embedding, $e_{i} \in R^{K}$ represents an embedding vector with dimension $K$.
    
    \subsubsection{Attention Module}
    Most of the advanced recommender systems use attention mechanism to capture user interests,  especially on the representation learning of user sequential behavior. We follow BST~\cite{chen2019behavior} and use multi-head self-attention~\cite{vaswani2017attention} to learn deep representation of user interests based on the \textit{User Behavior} features, which can be formulated as follows:
    \begin{equation}
        \label{for:multi-head attention}
        \text{MultiHead}(Q,K,V) = \text{Concat}(head_1,...,head_H) W^O \\
    \end{equation}
    \begin{equation}
        \begin{split}
            head_i &= \text{Attention}(QW_i^Q,KW_i^K,VW_i^V) \\
            &= \text{Softmax}(\frac{QW_i^{Q} (KW_i^K)^T}{\sqrt{d_k}})VW_i^V    \\
        \end{split}
    \end{equation}
    where $Q$, $K$, $V$ are embedding matrices of \textit{User Behavior} features, which are converted through linear projection. $H$ denotes the number of attention heads, $d$ is the last dimension of embedding, $W_i^Q, W_i^K, W_i^V, W^O$ are all linear projection matrices. 
    
    \subsection{Multi-layer Perceptron (MLP)}
    As most of the recent deep models in recommender systems, the output of \textit{Embedding Module} and \textit{Attention Module} are concatenated, then fed into MLP with fully connected layers for final prediction.  
    
    The widely used loss function in recommendation is negative log-likelihood function, which can be defined as follows:
    \begin{equation}
    \begin{aligned}
        \mathcal{L} &= -\frac{1}{N}\sum_{(x,y)\in D}(y \log p(x) + (1-y) \log (1-p(x))) 
        \label{formula:loss}
    \end{aligned}
    \end{equation}
    where $x$ is the training sample, $y \in \{0,1\}$ is the corresponding label, represents whether user clicks target item. $p( \cdot )$ is the predicted output of model.
    
    \begin{figure*}[ht] 
    \centering
    \includegraphics[scale=1,width=\textwidth]{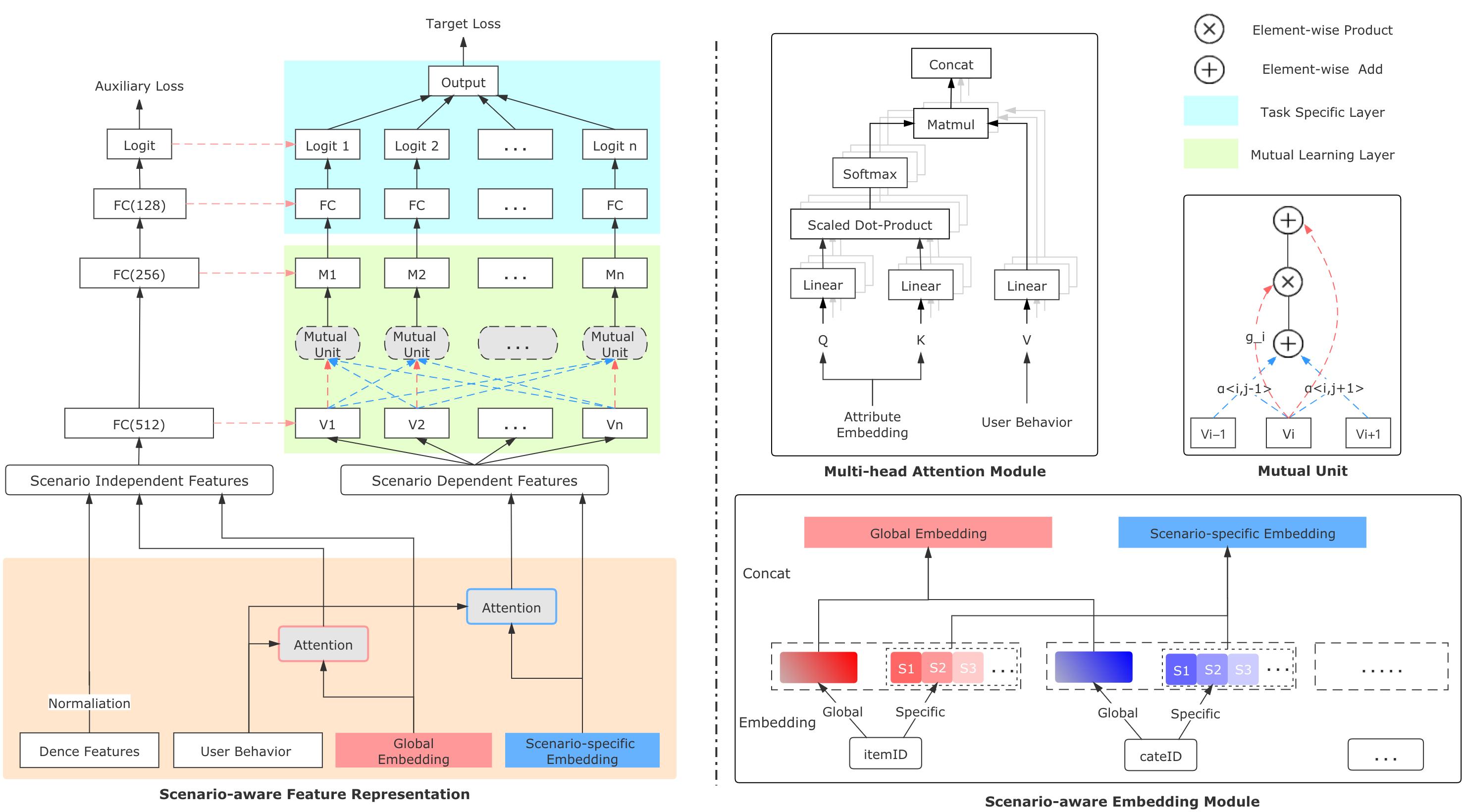}
    \caption{The overview of our proposed SAML model. The left part describes the model structure, the right part makes detail description of the modules involved. 
    From the bottom up, SAML has two main components: scenario-aware feature representation and scenario-mutual network. 
    In the first component, the raw features first go through the embedding module to obtain global and scenario-specific embedding vectors. Then part of features will enter the attention module to calculate the weights, and multiply the embedding of the user behavior to obtain the attention vector. 
    Finally, the feature vectors are concatenated and then fed into the two sub-networks of scenario-mutual network respectively. 
    The detail description of scenario-mutual network will be introduced in section \ref{sec:Scenario-mutual Network}.
    } 
    \label{fig:overall-HD}
    \end{figure*} 
    
    \subsection{Scenario-aware Feature Representation}
    Most existing recommendation methods mainly ignore the complex correlation between multiple scenarios, we first build both global and scenario-specific subspace for embedding and attention module, and then combine the features from  each subspace respectively, thus construct two types of feature which we named scenario-independent and scenario-dependent features.
    
    \subsubsection{Embedding Module}
    As depicted in Figure \ref{fig:overall-HD}, the embedding module explicitly embeds each feature into both global and scenario-specific subspace in parallel. And then combine the feature vectors from each subspace respectively to construct two types of embedding vectors.
    The motivation is to realize the perception of features to the global and specific scenarios. 
    As a comparison, that is more effective than directly increasing the embedding size, because no matter how to expand the dimension, the information between various scenarios is still mixed together without distinction. While our method has an explicit distinction between different scenarios. The ablation study in  section \ref{sec:Q4} also confirms this.
    
    \subsubsection{Attention Module}
    On top of the embedding module, the attention module is subsequently enriched to capture use interests both in global and specific scenarios. 
    According to formula \ref{for:multi-head attention}, we can define the formula here as: $\text{MultiHead}(Q_{g},K_{g},V_{g})$ and  $\text{MultiHead}(Q_{l},K_{l},V_{g})$, where subscript $g$ and $l$ indicate the source of embedding vector (i.g., global and scenario-specific). It is notable that we share the same embedding vector $V_{g}$ above but the attention weights are computed from each scenario separately, which bridges the universal and specific knowledge as well as leveraging the rich diversity of user behaviors in global.

\subsection{Scenario-mutual Network}
    \label{sec:Scenario-mutual Network}
    The scenario-mutual network contains two subnetworks: an auxiliary network and a multi-branch network, which are used to learn the scenario-independent and scenario-dependent features in parallel. In addition, a novel mutual unit is incorporated in the multi-branch network to model the complex correlations (e.g., differences and similarities) between multiple scenarios in an explicit manner. 
    Now we introduce these three components in detail.
    \subsubsection{Auxiliary Network}
    In MMoE, all experts learn the shared knowledge together, if a definite domain knowledge exists for each task, integrating it into the expert is not convenient in practice. 
    Therefore, we build the auxiliary network on top of scenario-independent features, which is used to learn shared knowledge from a global perspective. 
    Specifically, we not only obtain its final output (for supervised learning), but also extract its hidden layer representation and use as an additional input to the multi-branch network. 
    As shown in Figure~\ref{fig:overall-HD}, the knowledge propagates from auxiliary network to multi-branch network unidirectionally. 
    The advantage is that we can extract some universal and  scenario-independent knowledge to enhance the global perception of each specific scenario.
    An auxiliary loss function of negative log-likelihood is used to supervise the learning process.
    \subsubsection{Multi-branch Network}
    As shown in Figure~\ref{fig:overall-HD}, the input of each layer in multi-branch network contains two parts: output from the previous layer and hidden layer representation transferred from the auxiliary network. 
    Taking the $i$ th branch and $l$ th layer as example, the process can be formulated as
    \begin{equation}
        \begin{split}
            {V_a^l} &= \delta (V_a^{l-1}W_{a}^l + b_a^l) \\
            {V_{m_{i}}^l} &= \delta (\lbrack V_{m_{i}}^{l-1}, V_{a}^l \rbrack W_{m_{i}}^{l} +  b_{m_{i}}^l) \\ 
        \end{split} 
    \end{equation}
    where $ V_a^l $ is the $l$ th layer output from auxiliary network, $ V_{m_{i}}^l $ is the $l$ th layer output from $i$ th branch in multi-branch network,
    $W_{a}^l$, $b_a^l$, $W_{m_{i}}^{l}$, $b_{m_{i}}^l$ are the corresponding weight and bias, $\delta$ is the activation function.
    
    In order to make each branch clearly correspond to a specific scenario, and can only be optimized by the data of the scenario itself,
    we implement this by adding a mask on the connection between networks to stop the gradient back-propagation. Thus the gradient of instance $t$ which belongs to scenario $S_i$ will only back-propagate to update the parameters in branch $i$. 
    Finally, the total loss can be calculated as: 
    \begin{equation}
        \begin{split}
            \mathcal{L}_{total} &= \mathcal{L}_{target} + \mathcal{L}_{aux} \\
            &= \sum_{i}^{N} \mathcal{L}_{i}^{t} \cdot  {I}_{i}^{t} + \mathcal{L}_{aux} \\
            \textstyle{where} & \quad {I}_{i}^{t} = \left\{\begin{array}{lr}
                        1 \qquad \scriptstyle{if\ t\ \in S_{i}} \\
                        0 \qquad \scriptstyle{otherwise} \\
                \end{array} \right.
        \end{split} 
    \end{equation}
    where $\mathcal{L}_{target}$ and $\mathcal{L}_{aux}$ are the losses of multi-branch network and auxiliary network respectively, $N$ is the number of branches, $t$ is the training sample, $S_{i}$ is the data set of the $i$ th scenario, $\mathcal{L}_{i}^{t}$ is the loss of the $i$ th branch on sample $t$, $I(\cdot)$ is used as indicator function to constraint the gradient.
    
    \subsubsection{mutual unit}
    By combining the learned representation from auxiliary and multi-branch network, we can take advantage of both global and scenario-specific features. 
    However, the branches are independent of each other, which means that we actually isolate the relationship between scenarios and ignore the fact that similarity exists between part of scenarios to some extent. 
    In order to simultaneously model the differences and similarities between scenarios, we introduce a novel mutual unit that can enhance the representation learning by considering the similarity among multiple scenarios and alleviating the problem of insufficient training on part of scenarios as well.
    
    As shown in Figure~\ref{fig:overall-HD}, the mutual unit use the hidden layer output $V_i$ from $i$ th branch to calculate the cosine distance with other $V_{j,j \ne i}$ to capture the similarities between different scenarios. A light-weight gate network is designed to control the degree learning from other similar scenarios. The learning procedure can be defined as follows:
    \begin{equation}
        {M_i} = V_i + g_i *\sum_{j=1,j \ne i}^{N}(\alpha_{ij}*V_j) 
    \end{equation}
    \begin{equation}
        g_i = \text{Sigmoid}(W_i V_i + b_i)
    \end{equation}
    \begin{equation}
        \alpha_{ij} = \text{Softmax}(\frac{\cos{<V_i, V_j>}}{\sum_{j=1,j\ne i}^{N} {\cos{<V_i, V_j>}}})
    \end{equation}
    where $V_i \in \mathbb{R}^{D}$, $D$ is the dimension of the hidden layer output of each branch, 
    $g_i \in \mathbb{R}$ is the gate coefficient learned from $V_i$. 
    $\alpha_{ij} \in \mathbb{R}$ is the normalized similarity  coefficient indicating the similarity between scenario $i$ and $j$.
    $W_i \in \mathbb{R}^{D}$, $b_i \in \mathbb{R}$ are the linear matrix and bias of gate network. 
    
    Finally, $M_{i}$ of each branch are sent to the next layer respectively, which benefits from the assistance of the similar scenarios, having the advantages of: (1) it maintain the dominance of the current branch to the greatest extent, so it can accurately model the differences between scenarios; (2) it can leverage the knowledge from some of the similar scenarios to enhance itself adaptively and vice versa.
    Note that when gate coefficient $g$ equal to 0, the similarity between scenarios is not considered, thus the network degenerates into multi-branch network with independent branches.
    

\section{Experiments and analysis}
In this section, we present our experiments in detail, including datasets, competitors, experimental setup, evaluation metrics, and the corresponding analysis. 
The experiments are intended to answer the following questions:
\begin{itemize}
\item \textbf{Q1}: How does our proposed model SAML compare with state-of-the-art methods on the recommendation task?
\item \textbf{Q2}: Does SAML really help to improve the recommendation results on each  scenario?
\item \textbf{Q3}: How is the effectiveness of critical technical designs in SAML?
\item \textbf{Q4}: How do different experimental settings (i.g., embedding size, number of attention heads, etc.) influence the performance of SAML?
\item \textbf{Q5}: How does SAML provide effective recommendation results intuitively?

\end{itemize}

\subsection{Datasets}
We conduct experiments on public and industrial datasets respectively, both of them are collected from real-world e-commerce platforms. Table \ref{tab:dataset table} summarizes the statistics of datasets used in this paper.

\textbf{Public Dataset\footnote{\url{https://tianchi.aliyun.com/dataset/dataDetail?dataId=56}}.} It's a public dataset released by Alimama, an online advertising platform in China. The dataset consists of 8 days of ad display/click logs from 2017-05-06 to 2017-05-12. We use the first 7 days for training and the last day for testing, and set behavior sequence length to 15.
We filter the samples of which user profile is missing, and then divide the scenario according to \textit{City\_level}. 

\textbf{Industrial Dataset.} 
It's an industrial dataset collected from the online recommender system of AliExpress, a cross-border e-commerce platform. Logs from 2019-08-24 to 2019-08-30 are used for training and 2019-08-31 is for testing, users' recent 30 behaviors are also recorded in logs. 
In this dataset, more than two hundred countries are contained, the performance of each country varies greatly, thus we divide the scenario according to \textit{Country\_id}. 

\subsection{Competitors}
To evaluate the performance of proposed method, we compare SAML with the following models: 
\begin{itemize}
\item \textbf{Wide\&Deep}\cite{cheng2016wide} contains a wide part for memorization and a deep part for generalization, we implement the wide part by linear regression (LR) and the deep part by MLP.
\item \textbf{DeepFM}\cite{guo2017deepfm} replace LR with factorization machines (FM) and combine with MLP to model low- and high-order feature interactions. The two components share embedding space and summed up their outputs as the final prediction.
\item \textbf{DCN}\cite{wang2017deep} introduces a novel cross network that is more efficient in learning certain bounded-degree feature interactions. 
Our implementation follows the same structure as original paper and stacks only three cross layers.
\item \textbf{DIN}\cite{zhou2018deep} represents user interest with regard to the target item by adaptively learning the attention weight.
\item \textbf{MMoE}\cite{ma2018modeling} is a multi-task learning approach that use Multi-gate Mixture-of-Experts to model task relationships from data.
We implement each expert with a two-layer MLP to learn representation from multiple scenarios.
\item \textbf{BST}\cite{chen2019behavior} utilizes the powerful Transformer model to capture the sequential signals underlying users’ behavior sequences for recommendation. We regard it as a base model for comparison in this paper.

\end{itemize}

\subsection{Experimental Setup}
We implement our experiments on a distribution TensorFlow framework\footnote{\url{https://www.aliyun.com/product/bigdata/product/learn}}. All competitors in the experiments use ReLU activation function and Adam~\cite{kingma2014adam} optimizer. 
For each dataset, their settings are as follows: 

\textbf{Public settings.} 
Learning rate is tuned and set to be 5e-4, mini-batch size is set to 128. The hidden layer size of MLP involved are set by 128 $\times$ 64. The size of global embedding and scenario-specific embedding for each attribute is set to 12 and 4 respectively. The number of attention heads is set to 4. 
Corresponding to \textit{Country\_level} attribute, the scenarios are divided into 5 parts.

\textbf{Industrial settings.} 
Learning rate is tuned and set to be 1e-4, mini-batch size is set to 1024. The hidden layer size of MLP involved is set by 256 $\times$ 128. The size of global embedding and scenario-specific embedding for each attribute are set to 8 and 2 respectively. 
The number of attention heads is set to 4.
Corresponding to \textit{Country\_id} attribute, the traffic of top 9 countries exceeds sixty percent, thus we select top-9 countries with the highest traffic volume and group the rest into one.

\subsection{Evaluation Metrics.}
\textbf{AUC.}
We use AUC (Area under ROC Curve) as our metric for measurement of model performance. It is defined as:
\begin{equation}
    {AUC} = \frac{1}{|D^+||D^-|}\sum_{x^+ \in D^+} \sum_{x^- \in D^-}{I}(f(x^+)>f(x^-))
\end{equation}
where $D^+$ and $D^-$ donate the collection of positive and negative samples respectively, $|D^+|$ and $|D^-|$ donate the number of samples in $D^+$ and $D^-$, $f(\cdot)$ is the output of model, $I(\cdot)$ is the indicator function.

\textbf{RelaImpr.}
We follow~\cite{yan2014coupled} to introduce RelaImpr metric to measure relative improvement over models. For a random guesser, the value of AUC is 0.5. Hence RelaImpr is defined as below:
\begin{equation}
{RelaImpr} = \left(\frac{AUC(measured~model)-0.5}{AUC(base~model)-0.5}-1 \right) \times 100\%
\end{equation}


\begin{table}
    \centering
  \caption{STATISTICS OF PUBLIC AND INDUSTRIAL DATASETS}
  \label{tab:dataset table}
  \begin{tabular}{c r r r r r}
    \toprule
    Dataset      & User  & Item    & Click   & Conversion  & Samples \\
    \midrule
    Public       & 1.32M  & 1.08M   & 1.23M    & -           & 24.6M  \\
    Industrial   & 35.9M   &111M     & 307M    & 1.90M        & 2.85B \\
  \bottomrule
\end{tabular}
\end{table}


\subsection{Overall Performance (Q1)}
We conduct experiments of multi-scenario recommendation on both public and industrial datasets. 
The corresponding results are present in table \ref{tab:CTRCVR on AEAM}, which are recorded from comparison models learning through the whole scenarios, and then testing the overall performance. 

From above results, we have several important observations: 
(1) DIN performs better than previous models, mainly attributed to the attention mechanism which captures the user interest with regard to target item; 
(2) MMoE performs better than DIN and previous models, especially on industrial dataset of which expert networks can implicitly model correlations between scenarios, indicating that discrepancy between scenarios can not be neglected;
(3) BST is a strong competitor, and is just slightly worse than MMoE, indicating the effectiveness of incorporating Transformer into the recommendation model;
(4) SAML achieves the best performance on both datasets, demonstrates its effectiveness and generalization capability on multiple scenarios. Note that on industrial dataset, SAML achieves 0.0096 absolute AUC gain over BST, which is a significant improvement for business.

\begin{table}
    \centering
  \caption {MODEL COMPARISON ON PUBLIC AND INDUSTRIAL DATASETS}
  \label{tab:CTRCVR on AEAM}
  \begin{tabular}{c c c c c}
    \toprule
    \multirow{2}{*}{Model} & \multicolumn{2}{c}{Public} & \multicolumn{2}{c}{Industrial}  \\
     & AUC & RelaImpr$^{\mathrm{a}}$ & AUC & RelaImpr$^{\mathrm{a}}$ \\
    \midrule
    W\&D          & 0.6320  & -2.07\%         & 0.7245         & -7.61\% \\
    DeepFM        & 0.6329  & -1.40\%         & 0.7308         & -5.02\%   \\
    DCN           & 0.6333  & -1.11\%         & 0.7308         & -5.02\% \\
    DIN           & 0.6343  & -0.37\%         & 0.7409         & -0.86\%  \\
    MMoE          & 0.6357   & 0.66\%         & 0.7449         & 0.78\%   \\
    BST           & 0.6348    & 0.00\%        & 0.7430         & 0.00\%   \\
    SAML       & \textbf{0.6392} &  \textbf{3.26}\%   & \textbf{0.7526}   & \textbf{3.95\%}   \\
    \bottomrule
    \multicolumn{4}{l}{$^{\mathrm{a}}$RelaImpr is based on BST.}
    \end{tabular}
\end{table}


\subsection{Single Scenario Performance (Q2)}
To verify whether our proposed SAML can really help to improve the recommendation results in each scenario, we take BST and SAML as comparison, and conduct experiments on industrial dataset according to the following procedure: 
(1) BST-Individual trains with the data of each scenario individually; (2) BST trains with the data combined from all scenarios; (3) SAML trains with the data combined from all scenarios and incorporates the critical technical designs we proposed above. And then all the models are tested on each scenario separately.

Table \ref{tab:CTR on each country} shows the comparison results of several models in each scenario. According to the results, we have several observations: (1) By learning from multiple scenarios, BST is better than BST-Individual, indicating that the information between different scenarios can be mutually used to promote each other; (2) SAML consistently outperforms BST on each scenario, demonstrating the effectiveness of our scenario-aware mutual learning approach, which explicitly consider the differences and similarities between scenarios. 

\begin{table}
    \centering
  \caption {SINGLE SCENARIO COMPARISON (AUC) ON INDUSTRIAL DATASET}
  \label{tab:CTR on each country}
  \begin{tabular}{c c c c}
    \toprule
    Scenario & BST-Individual & BST & SAML \\
    \midrule
    RU          & 0.7391         & 0.7475    & 0.7574\\
    BR          & 0.7399         & 0.7572    & 0.7683\\
    ES          & 0.7261         & 0.7499    & 0.7600\\
    US          & 0.7128         & 0.7407    & 0.7517\\
    FR          & 0.7219         & 0.7502    & 0.7592\\
    PL          & 0.7199         & 0.7496    & 0.7608\\
    NL          & 0.7083         & 0.7403    & 0.7520\\
    CL          & 0.7107         & 0.7466    & 0.7550\\
    UA          & 0.7051         & 0.7429    & 0.7500\\
    Others      & 0.7356         & 0.7421    & 0.7523\\
    \bottomrule
\end{tabular}
\end{table}

\begin{figure*}[htbp]
\centering
\label{tab:ablation study experiment setting}
\subfigure[embedding size]{
\begin{minipage}[bt]{0.33\linewidth}
\centering
\includegraphics[width=2.3in]{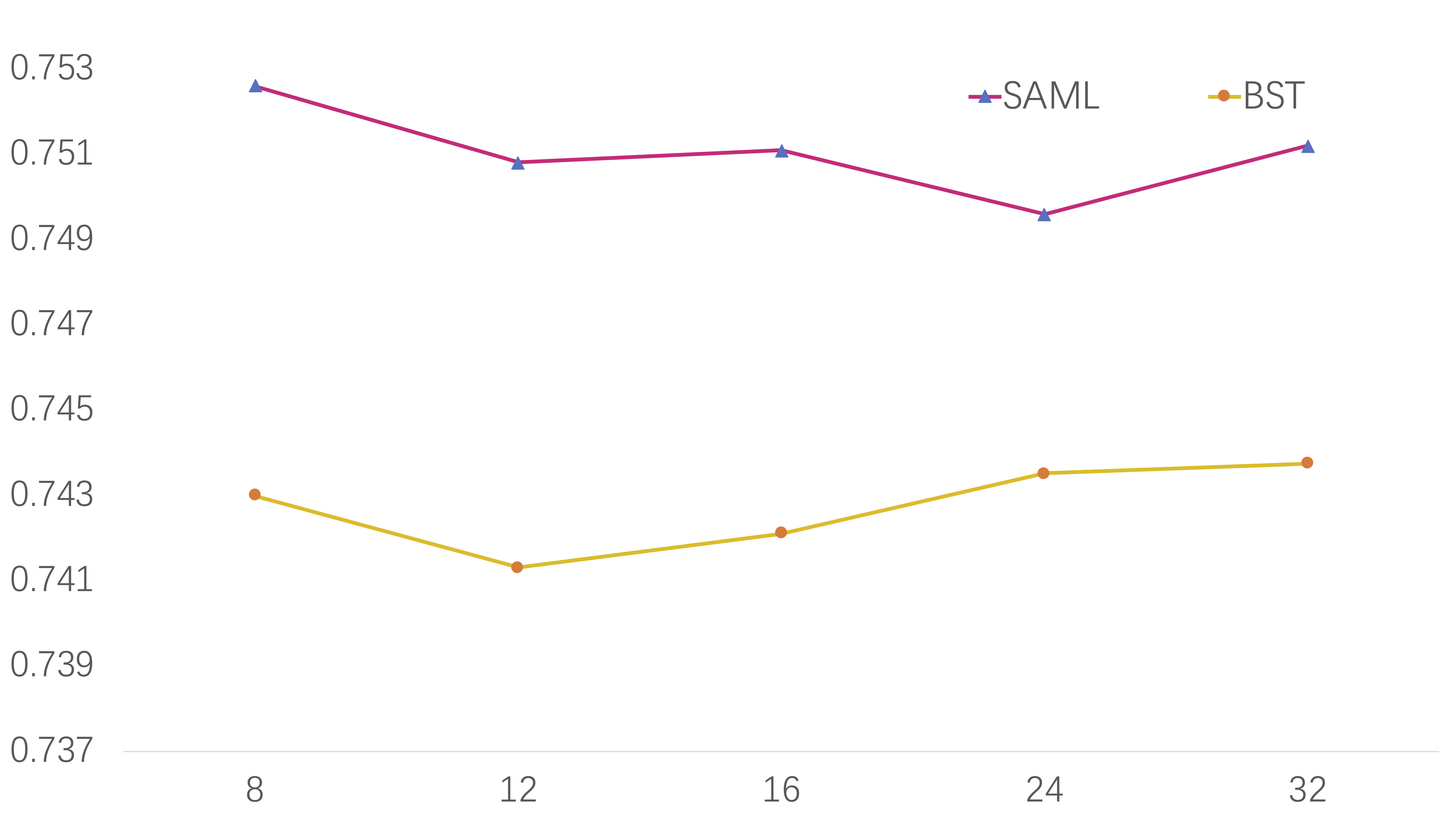}
\label{fig:embedding size}
\end{minipage}%
}%
\subfigure[attention heads]{
\begin{minipage}[bt]{0.33\linewidth}
\centering
\includegraphics[width=2.3in]{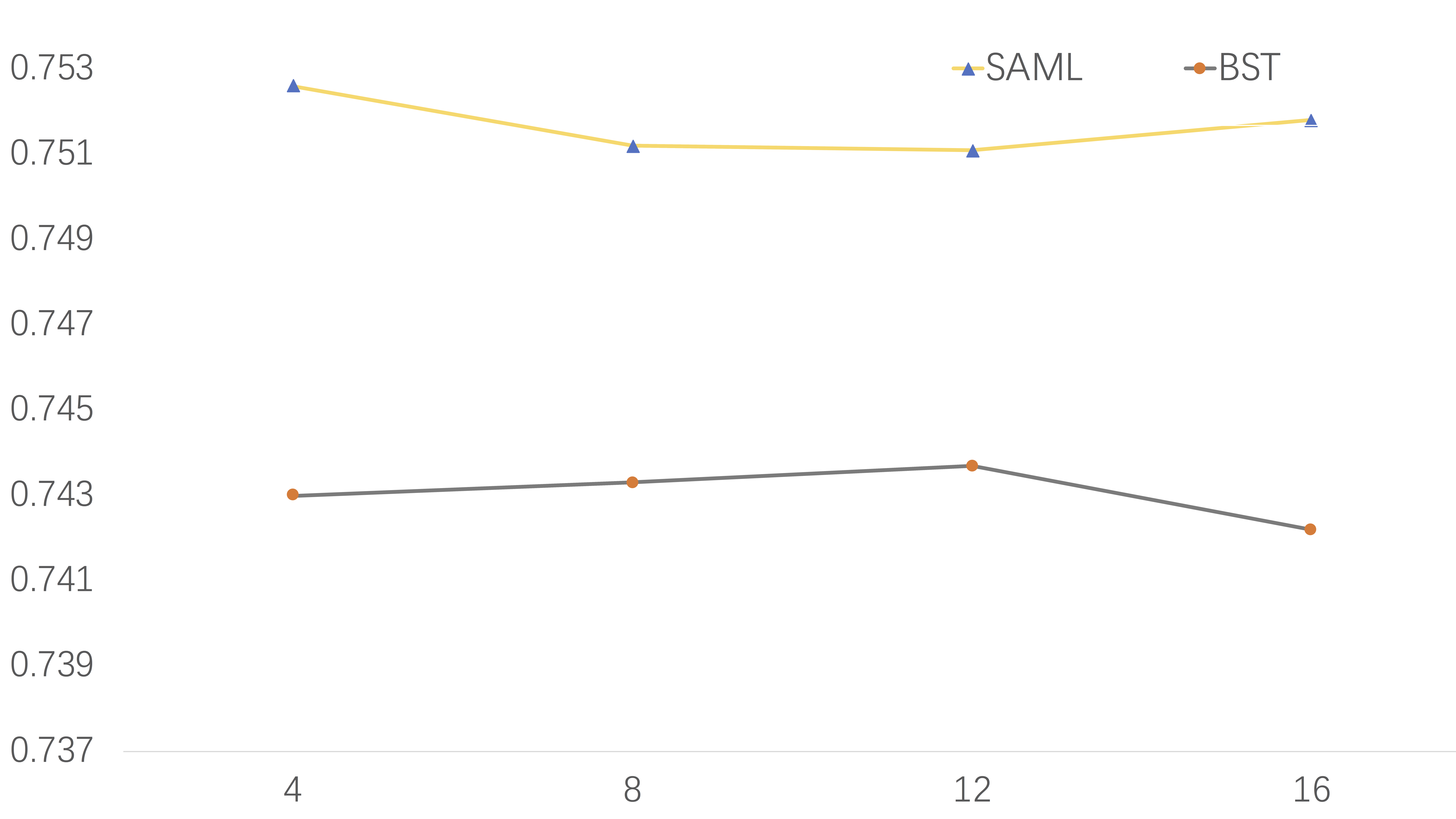}
\label{fig:number of attention heads}
\end{minipage}%
}%
\subfigure[depth of network]{
\begin{minipage}[bt]{0.33\linewidth}
\centering
\includegraphics[width=2.2in]{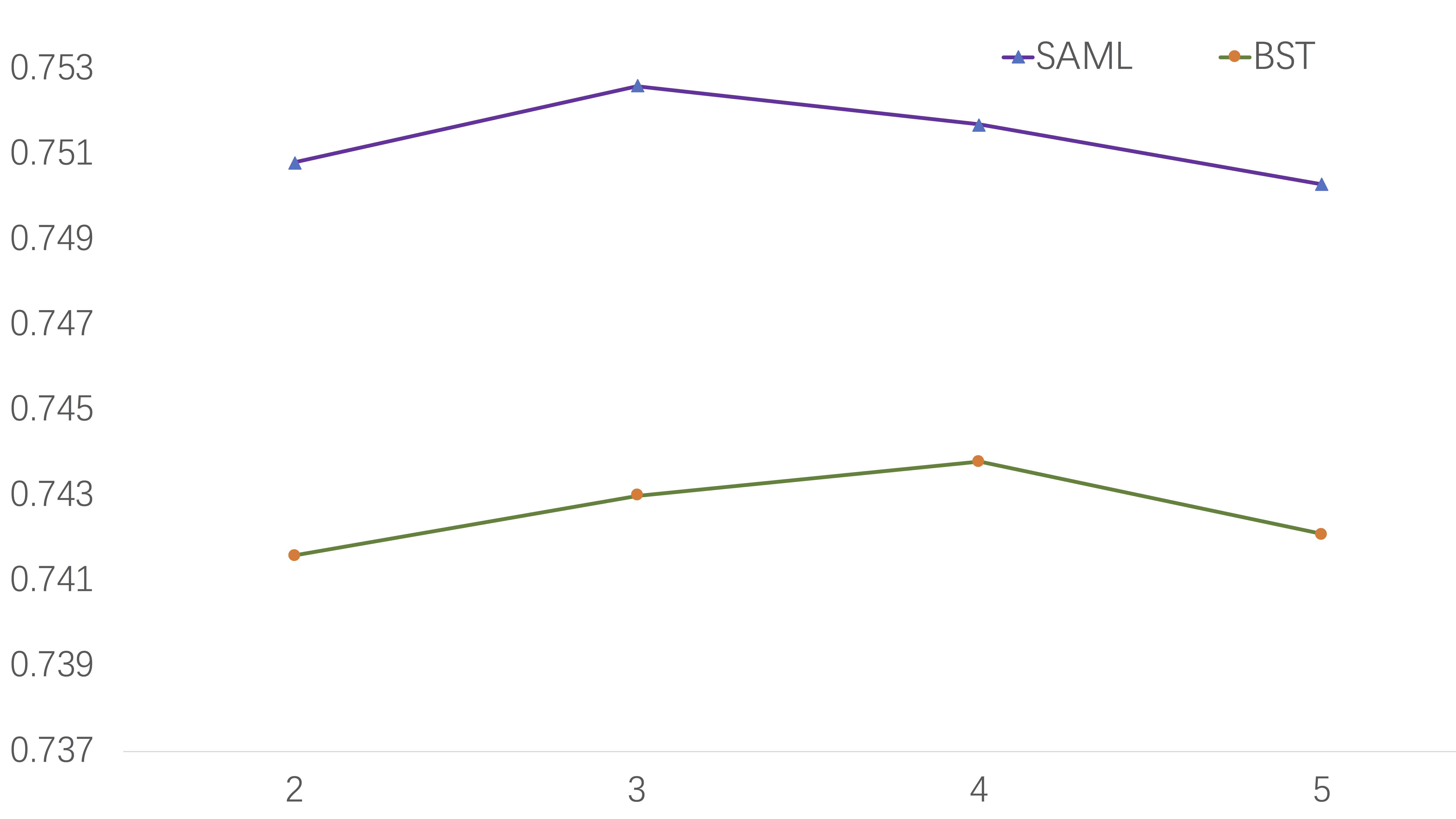}
\label{fig:depth of network}
\end{minipage}
}%
\centering
\caption{The results of different experimental settings in BST and SAML.}
\label{fig:ablation study}
\end{figure*}

\begin{table}
    \centering
  \caption {ABLATION RESULTS OF VARIANT MODELS ON INDUSTRIAL DATASET}
  \label{tab:ablation study}
  \begin{tabular}{c c c}
    \toprule
    Model & AUC & RelaImpr$^{\mathrm{a}}$ \\
    \midrule
    SAML                & \textbf{0.7526}   & \textbf{3.95\%} \\
    SAML w/o gate       & 0.7489            & 2.42\%  \\
    SAML w/o aux        & 0.7472            & 1.72\% \\
    SAML w/o gate\&mut  & 0.7457            & 1.11\%  \\
    BST                 & 0.7430            & 0.00\%  \\
    \bottomrule
\multicolumn{2}{l}{$^{\mathrm{a}}$RelaImpr is based on BST.}
\end{tabular}
\end{table}

\subsection{Ablation Study of Critical Technical Designs (Q3)}
To further verify the effectiveness of critical technical designs in SAML, such as the effectiveness of learning scenario-aware feature representation and the benefits of modeling differences and similarities between scenarios, we conduct ablation experiments to compare SAML with the following variant models:  
\begin{itemize}
\item {SAML w/o gate}: removes the mutual unit between scenarios, equal to fix the gate coefficient $g$ as 0, which means that it will not consider the similarity between scenarios.
\item {SAML w/o aux}: removes the auxiliary network upon scenario-independent features, combine the two features together then fed into mutual network, thus the differences between scenarios are not fully considered.
\item {SAML w/o gate\&mut}: removes the mutual network upon scenario-dependent features, thus the features are combined and fed into auxiliary network (a unify MLP), which means that it considers neither similarities nor differences between scenarios.
\end{itemize}

Table \ref{tab:ablation study} shows the performance of SAML and different variants, the state-of-the-art recommendation model BST is also included as a base model. Based on the experiment results, we have the following observations:

\begin{itemize}
\item {SAML w/o gate} performs worse than SAML about 0.0037 absolute AUC, indicating the effectiveness of mutual unit which can automatically select similar representations from other scenarios for enhancement.
\item {SAML w/o aux} declines by about 0.0054 absolute AUC than SAML, demonstrating the effectiveness of separation of scenario-specific representation learning from the global representation learning.
\item From the comparison between SAML w/o gate\&mut and BST, we can get a rough view of the contribution from scenario-aware feature representation, which achieves 0.0024 absolute AUC improvement than base model.
\end{itemize}

\subsection{Ablation Study of Experimental Settings (Q4)}
\label{sec:Q4}
In addition to the ablation study of critical technical designs in SAML, we also study the sensitivity of the model to different experimental settings includes embedding size, number of attention heads, and depth of network layers. 

The embedding and attention module in scenario-aware feature representation are different from directly increasing the embedding size or number of attention heads in that the feature diversity between scenarios are explicitly considered. 
However, one may question that the reason for the improvement we obtained may be purely due to the increase in embedding size or attention heads rather than the perception of scenario. 
Therefore, we compared the two models of BST and SAML, first fixed other settings, and then tune the embedding size and attention heads from 8, 12, 16, 24, 32 and 4, 8, 12, 16, respectively.
Similarly, we also compared different depths of network layers, including 2, 3, 4, and 5, respectively.

From the results in Figure~\ref{fig:embedding size} and Figure~\ref{fig:number of attention heads}, we can see that SAML model incorporates with scenario awareness consistently outperform the comparison model in various setting of embedding size and attention heads, while the improvement of purely increasing the embedding size or attention heads are all very limited. We consider that since increasing the embedding size or attention heads do have improvement to some extent, while the lack of scenario awareness may limit the expression capability of each attribute, hence resulting in limited improvement. 

Increasing the depth of network layers can enhance the model capacity but also potentially leads to over-fitting. As can be seen from Figure~\ref{fig:depth of network}, at the beginning stage, i.g., from two layers to four layers, increasing the number of hidden layers consistently improves the model’s performance. However, it saturates at five layers that increasing more layers even marginally decreases the AUC scores, where the model may overfit the training set. Therefore, we use three/two hidden layers for SAML in industrial/public dataset.

\begin{figure}[ht] 
\centering
\includegraphics[width=\linewidth]{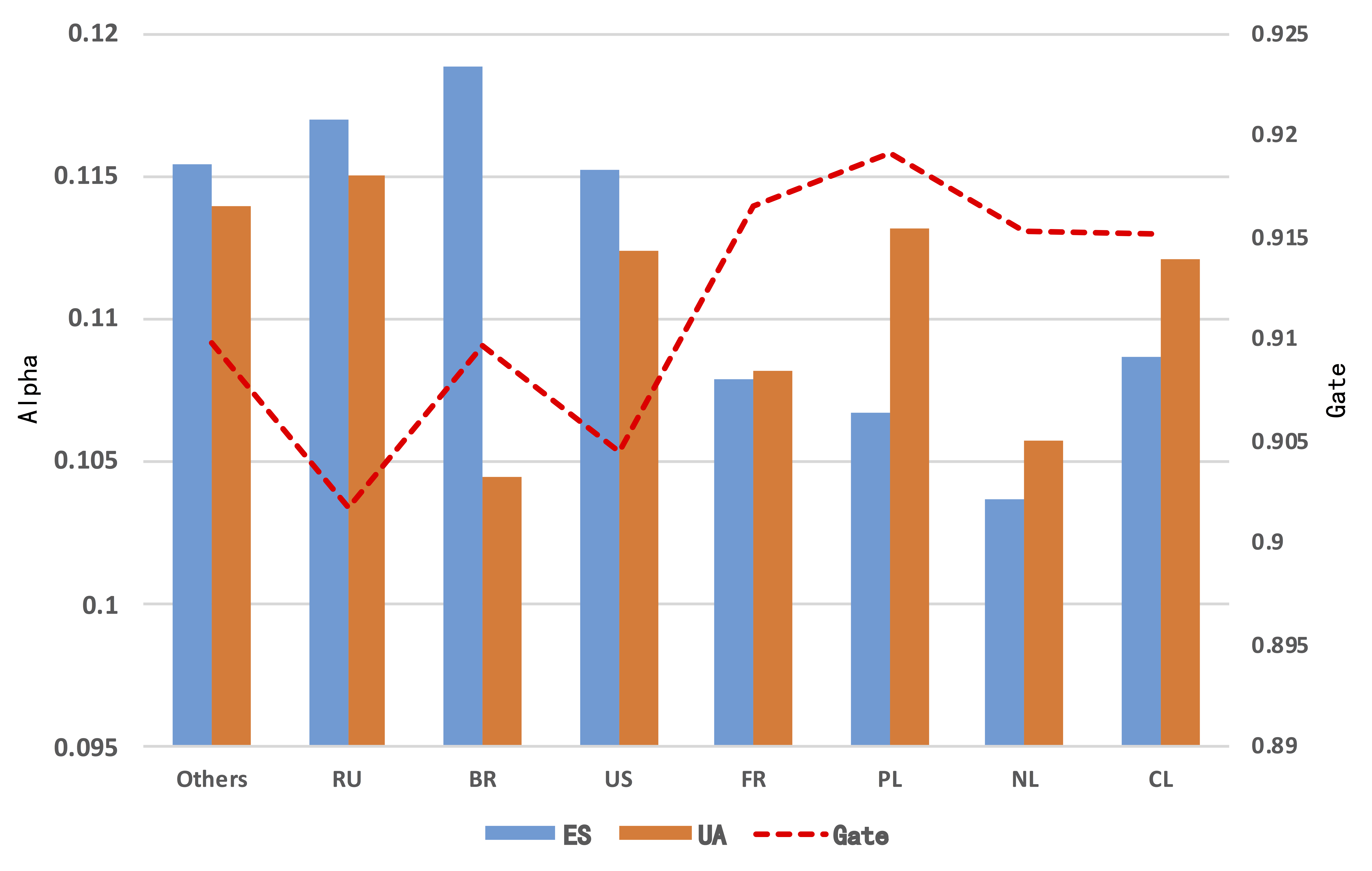}
\caption{Visualization of $\alpha$ and $g$ in mutual unit.} 
\label{fig:Visualalpha}
\end{figure}

\subsection{Visualization Analysis (Q5)}
The key to understanding how SAML provides effective recommendation results is to understand how mutual units help optimizing scenario correlation and enhance representation learning for each scenario.  
Thus we conduct experiments on industrial dataset to visualize the accumulated probability of $\alpha$ and $g$ in mutual units.
The similarity coefficient $\alpha$ is used to learn the similarity between scenarios and the gate coefficient $g$ is used to control the degree of learning from other similar scenarios, which is based on the learning situation of each scenario itself.

Taking Spain (ES) and Ukraine (UA) as example, the distributions of $\alpha$ are dramatically different as shown in Figure~\ref{fig:Visualalpha}. The relevance of ES to Brazil (BR) is much higher than that of UA because of the sport-related categories. In contrast, the similarity of UA with Poland (PL) is higher than that of ES for the geographical location. What's more, both of them are similar to Russia (RU), United States (US) and Others because these three scenarios have relatively larger volumes of traffic to affect model learning. 

The gate coefficient $g$ and the similarity coefficient $\alpha$ show a certain correlation, which further validates our assumptions: 
(1) scenarios with more traffic such as RU and BR can learn better representation through themselves thus tend to suppress the $g$ to reduce the effect from other scenarios; 
(2) scenarios with low traffic such as FR, NL, etc., due to insufficient learning of their own representations, thus tend to increase the $g$ to rely more on the representations of other similar scenarios.

\section{Conclusion}
In this paper, a novel recommendation model named Scenario-aware Mutual Learning (SAML) is proposed in the field of e-commerce recommendation to capture the complex correlations (e.g., differences and similarities) between multiple scenarios. 
First, we introduce scenario-aware feature representation to learn feature representations both in global and scenario-specific. 
Then we introduce an auxiliary network to model the shared knowledge across all scenarios, and use a multi-branch  network to model the differences among specific scenarios. 
Finally, we employ a mutual unit to adaptively learn the similarity of user's interests between various scenarios. 
An extensive set of experiments are provided to show the competitive performance of SAML and transferability of the learning framework. Detailed discussion of ablation studies and visualization analysis are also provided to show the insight of how the SAML works in real-world datasets.


\bibliographystyle{./bibliography/IEEEtran}
\bibliography{./bibliography/IEEEexample}

\begin{thebibliography}{10}
\providecommand{\url}[1]{#1}
\csname url@samestyle\endcsname
\providecommand{\newblock}{\relax}
\providecommand{\bibinfo}[2]{#2}
\providecommand{\BIBentrySTDinterwordspacing}{\spaceskip=0pt\relax}
\providecommand{\BIBentryALTinterwordstretchfactor}{4}
\providecommand{\BIBentryALTinterwordspacing}{\spaceskip=\fontdimen2\font plus
\BIBentryALTinterwordstretchfactor\fontdimen3\font minus
  \fontdimen4\font\relax}
\providecommand{\BIBforeignlanguage}[2]{{%
\expandafter\ifx\csname l@#1\endcsname\relax
\typeout{** WARNING: IEEEtran.bst: No hyphenation pattern has been}%
\typeout{** loaded for the language `#1'. Using the pattern for}%
\typeout{** the default language instead.}%
\else
\language=\csname l@#1\endcsname
\fi
#2}}
\providecommand{\BIBdecl}{\relax}
\BIBdecl

\bibitem{zhang2017joint}
Y.~Zhang, Q.~Ai, X.~Chen, and W.~B. Croft, ``Joint representation learning for
  top-n recommendation with heterogeneous information sources,'' in
  \emph{Proceedings of the 2017 ACM on Conference on Information and Knowledge
  Management}.\hskip 1em plus 0.5em minus 0.4em\relax ACM, 2017, pp.
  1449--1458.

\bibitem{lian2018towards}
J.~Lian, F.~Zhang, X.~Xie, and G.~Sun, ``Towards better representation learning
  for personalized news recommendation: a multi-channel deep fusion approach.''
  in \emph{IJCAI}, 2018, pp. 3805--3811.

\bibitem{zhou2018deep}
G.~Zhou, X.~Zhu, C.~Song, Y.~Fan, H.~Zhu, X.~Ma, Y.~Yan, J.~Jin, H.~Li, and
  K.~Gai, ``Deep interest network for click-through rate prediction,'' in
  \emph{Proceedings of the 24th ACM SIGKDD International Conference on
  Knowledge Discovery \& Data Mining}.\hskip 1em plus 0.5em minus 0.4em\relax
  ACM, 2018, pp. 1059--1068.

\bibitem{zhou2019deep}
G.~Zhou, N.~Mou, Y.~Fan, Q.~Pi, W.~Bian, C.~Zhou, X.~Zhu, and K.~Gai, ``Deep
  interest evolution network for click-through rate prediction,'' in
  \emph{Proceedings of the AAAI Conference on Artificial Intelligence},
  vol.~33, 2019, pp. 5941--5948.

\bibitem{feng2019deep}
Y.~Feng, F.~Lv, W.~Shen, M.~Wang, F.~Sun, Y.~Zhu, and K.~Yang, ``Deep session
  interest network for click-through rate prediction,'' \emph{arXiv preprint
  arXiv:1905.06482}, 2019.

\bibitem{chen2019behavior}
Q.~Chen, H.~Zhao, W.~Li, P.~Huang, and W.~Ou, ``Behavior sequence transformer
  for e-commerce recommendation in alibaba,'' in \emph{Proceedings of the 1st
  International Workshop on Deep Learning Practice for High-Dimensional Sparse
  Data}, 2019, pp. 1--4.

\bibitem{he2014practical}
X.~He, J.~Pan, O.~Jin, T.~Xu, B.~Liu, T.~Xu, Y.~Shi, A.~Atallah, R.~Herbrich,
  S.~Bowers \emph{et~al.}, ``Practical lessons from predicting clicks on ads at
  facebook,'' in \emph{Proceedings of the Eighth International Workshop on Data
  Mining for Online Advertising}.\hskip 1em plus 0.5em minus 0.4em\relax ACM,
  2014, pp. 1--9.

\bibitem{covington2016deep}
P.~Covington, J.~Adams, and E.~Sargin, ``Deep neural networks for youtube
  recommendations,'' in \emph{Proceedings of the 10th ACM conference on
  recommender systems}.\hskip 1em plus 0.5em minus 0.4em\relax ACM, 2016, pp.
  191--198.

\bibitem{borisyuk2017lijar}
F.~Borisyuk, L.~Zhang, and K.~Kenthapadi, ``Lijar: A system for job application
  redistribution towards efficient career marketplace,'' in \emph{Proceedings
  of the 23rd ACM SIGKDD International Conference on Knowledge Discovery and
  Data Mining}, 2017, pp. 1397--1406.

\bibitem{ma2018modeling}
J.~Ma, Z.~Zhao, X.~Yi, J.~Chen, L.~Hong, and E.~H. Chi, ``Modeling task
  relationships in multi-task learning with multi-gate mixture-of-experts,'' in
  \emph{Proceedings of the 24th ACM SIGKDD International Conference on
  Knowledge Discovery \& Data Mining}.\hskip 1em plus 0.5em minus 0.4em\relax
  ACM, 2018, pp. 1930--1939.

\bibitem{cheng2016wide}
H.-T. Cheng, L.~Koc, J.~Harmsen, T.~Shaked, T.~Chandra, H.~Aradhye,
  G.~Anderson, G.~Corrado, W.~Chai, M.~Ispir \emph{et~al.}, ``Wide \& deep
  learning for recommender systems,'' in \emph{Proceedings of the 1st workshop
  on deep learning for recommender systems}.\hskip 1em plus 0.5em minus
  0.4em\relax ACM, 2016, pp. 7--10.

\bibitem{guo2017deepfm}
H.~Guo, R.~Tang, Y.~Ye, Z.~Li, and X.~He, ``Deepfm: a factorization-machine
  based neural network for ctr prediction,'' \emph{arXiv preprint
  arXiv:1703.04247}, 2017.

\bibitem{wang2017deep}
R.~Wang, B.~Fu, G.~Fu, and M.~Wang, ``Deep \& cross network for ad click
  predictions,'' in \emph{Proceedings of the ADKDD'17}.\hskip 1em plus 0.5em
  minus 0.4em\relax ACM, 2017, p.~12.

\bibitem{he2016deep}
K.~He, X.~Zhang, S.~Ren, and J.~Sun, ``Deep residual learning for image
  recognition,'' in \emph{Proceedings of the IEEE conference on computer vision
  and pattern recognition}, 2016, pp. 770--778.

\bibitem{vaswani2017attention}
A.~Vaswani, N.~Shazeer, N.~Parmar, J.~Uszkoreit, L.~Jones, A.~N. Gomez,
  {\L}.~Kaiser, and I.~Polosukhin, ``Attention is all you need,'' in
  \emph{Advances in neural information processing systems}, 2017, pp.
  5998--6008.

\bibitem{caruana1997multitask}
R.~Caruana, ``Multitask learning,'' \emph{Machine learning}, vol.~28, no.~1,
  pp. 41--75, 1997.

\bibitem{argyriou2008spectral}
A.~Argyriou, M.~Pontil, Y.~Ying, and C.~A. Micchelli, ``A spectral
  regularization framework for multi-task structure learning,'' in
  \emph{Advances in neural information processing systems}, 2008, pp. 25--32.

\bibitem{ni2018perceive}
Y.~Ni, D.~Ou, S.~Liu, X.~Li, W.~Ou, A.~Zeng, and L.~Si, ``Perceive your users
  in depth: Learning universal user representations from multiple e-commerce
  tasks,'' in \emph{Proceedings of the 24th ACM SIGKDD International Conference
  on Knowledge Discovery \& Data Mining}.\hskip 1em plus 0.5em minus
  0.4em\relax ACM, 2018, pp. 596--605.

\bibitem{ma2018entire}
X.~Ma, L.~Zhao, G.~Huang, Z.~Wang, Z.~Hu, X.~Zhu, and K.~Gai, ``Entire space
  multi-task model: An effective approach for estimating post-click conversion
  rate,'' in \emph{The 41st International ACM SIGIR Conference on Research \&
  Development in Information Retrieval}.\hskip 1em plus 0.5em minus 0.4em\relax
  ACM, 2018, pp. 1137--1140.

\bibitem{wen2019conversion}
H.~Wen, J.~Zhang, Y.~Wang, W.~Bao, Q.~Lin, and K.~Yang, ``Conversion rate
  prediction via post-click behaviour modeling,'' \emph{arXiv preprint
  arXiv:1910.07099}, 2019.

\bibitem{jacobs1991adaptive}
R.~A. Jacobs, M.~I. Jordan, S.~J. Nowlan, G.~E. Hinton \emph{et~al.},
  ``Adaptive mixtures of local experts.'' \emph{Neural computation}, vol.~3,
  no.~1, pp. 79--87, 1991.

\bibitem{zhang2018deep}
Y.~Zhang, T.~Xiang, T.~M. Hospedales, and H.~Lu, ``Deep mutual learning,'' in
  \emph{Proceedings of the IEEE Conference on Computer Vision and Pattern
  Recognition}, 2018, pp. 4320--4328.

\bibitem{hinton2015distilling}
G.~Hinton, O.~Vinyals, and J.~Dean, ``Distilling the knowledge in a neural
  network,'' \emph{arXiv preprint arXiv:1503.02531}, 2015.

\bibitem{kingma2014adam}
D.~P. Kingma and J.~Ba, ``Adam: A method for stochastic optimization,''
  \emph{arXiv preprint arXiv:1412.6980}, 2014.

\bibitem{yan2014coupled}
L.~Yan, W.-J. Li, G.-R. Xue, and D.~Han, ``Coupled group lasso for web-scale
  ctr prediction in display advertising,'' in \emph{International Conference on
  Machine Learning}, 2014, pp. 802--810.

\end{thebibliography}

\vspace{12pt}

\end{document}